# Single-Chip 1.024 Tb/s Optical Receiver for High-Speed Optical links

Ali Pirmoradi[1,†], Zhehao Yu[1,†], Amirreza Shoobi[1], Geun Ho Ahn[2], Jelena Vučković[2], and Firooz Aflatouni[1,*]

[1]Department of Electrical and Systems Engineering, University of Pennsylvania, Philadelphia, PA, USA

[2]E. L. Ginzton Laboratory, Stanford University, Stanford, CA, USA.

[*]firooz@seas.upenn.edu

[†]Authors with equal contributions

**Integrated optical transceivers, utilizing wavelength-division-multiplexing, offer a path forward for implementation of compact, high-bandwidth and energy-efficient interconnects for future data centers. Here we report the demonstration of a monolithically integrated optical receiver in 45nm CMOS, where efficient multi-layer optical demultiplexing with capacitive tuning control, energy-efficient electronics and wideband inverse designed grating couplers enable implementation of a 32-channel receiver chip based on wavelength-division multiplexing. The chip operates at an aggregate data-rate of 1.024 Tb/s with all channels operating simultaneously at a data-rate of 32 Gb/s/channel achieving a record energy efficiency of 71 fJ/b, including the power consumption of both the electronic circuitry and the tuning and control of photonic devices, and a record bandwidth density of 4 Tb/s/mm$^2$. The receiver achieves a bit-error-rate below $10^{-12}$ without requiring equalization, error correction or digital processing. Inverse-designed broadband grating couplers provide efficient, low-loss optical coupling into the chip. An on-chip demultiplexer, composed of Mach-Zehnder interferometers (MZIs) and ring resonators, offers a large channel-to-channel isolation sufficient for error-free operation. Capacitive phase shifters embedded within the ring resonators of the demultiplexer are used for wavelength alignment at a zero static power consumption. MZIs and ring-resonators are periodically selected and autonomously locked to the wavelength of the corresponding optical carrier. The implemented monolithic receiver offers a scalable, energy-efficient and reliable solution for the beyond Tb/s optical interconnects.**

The growing demand of computing, particularly driven by the growth of artificial intelligence (AI), machine learning (ML), and other high-performance computing (HPC) workloads are reshaping the data center infrastructure, calling for unprecedented performance scaling. Meanwhile, the energy consumption of hyperscale data centers continues to increase, further stressing local power grids, with the interconnect energy consumption becoming a significant contributor to their overall power budgets. Optical interconnects offer energy efficiency, bandwidth density and a high data-rate and have become a promising path to address the emerging HPC workload requirements.

Silicon photonics has been instrumental in advancing high-speed optical I/O technologies. Unlike free-space optics that mainly relies on discrete components, integrated photonics enables a significant reduction in system footprint, leading to enhanced performance and higher energy efficiency. Moreover, integration of on-chip transceiver (TRX) modules facilitates utilization of multiplexing schemes such as wavelength-division multiplexing (WDM), which significantly increases the overall link bandwidth by transmitting multiple optical data channels through a single fiber[1-11].

An integrated optical transmitter or receiver system can be implemented either through hybrid (co-packaged) or monolithic (on-the-same-chip) approaches. In hybrid integration, two separate chips, the electronic integrated circuit (EIC) and the photonic integrated circuit (PIC), are typically connected via flip-chip bonding or through an interposer[6-8,10]. While hybrid integration offers flexibility in selecting electronic and photonic fabrication process technologies, the parasitic components introduced by the complex electronic-photonic packaging between the PIC and EIC can limit energy efficiency and operational bandwidth. Furthermore, as the number of channels increases, the footprint of EIC-PIC interconnect pads used for tuning, control and drive of photonic devices within the PIC becomes a dominant factor in system scalability. In contrast, monolithic integration minimizes parasitic effects and packaging complexity, leading to an enhanced energy efficiency, higher per-channel data-rates, and greater areal bandwidth density[4,9,12]. While monolithic WDM systems offer certain advantages, the close integration of electronic and photonic components on a shared substrate often introduces thermal crosstalk, which can degrade the performance of temperature-sensitive photonic devices.

To achieve a large aggregate data-rate while maintaining low energy consumption and high areal bandwidth density, the design of electronics favors a moderate per-channel data-rate and an increased number of parallel WDM channels, which should be balanced with the required bandwidth of photonic devices and complexity of multi-wavelength light sources.

A WDM receiver typically consists of an optical demultiplexer (DeMux) followed by an array of photo-receivers. Optical DeMux systems are critical components that directly affect key performance metrics such as bit-error rate (BER) and energy efficiency. Achieving ultra-low power consumption and error-free operation at data-rates beyond 1 Tb/s requires a DeMux design with a low insertion loss and a high channel-to-channel isolation. Furthermore, such a device should be robust against environmental fluctuations and fabrication process variations to avoid unwanted resonance shifts and phase and amplitude distortions in the optical signal[13]. Athermal design strategies can reduce dependence on active thermal tuning[14], however, the effect of device mismatches caused by process variations[15,16] is typically not addressed in such designs. Photonic Mux/DeMux architectures based on thermally tuned Mach-Zehnder interferometers (MZIs) or ring resonators, have been demonstrated[17-20]. While thermal tuning provides a relatively large tuning range, it typically suffers from a high power consumption and increased complexity of multi-channel device tuning due to thermal crosstalk[21]. Monolithic integration of photonic devices with CMOS electronics enables the repurposing of the MOSFET gate structure to implement semiconductor-insulator-semiconductor capacitor based optical phase shifters that can perform tuning of photonic devices at a zero static power consumption[4,22].

In a typical WDM-based optical receiver, the outputs of optical DeMux are photo-detected using an array of photodiodes and the resulting photocurrents are amplified using a trans-impedance amplifier (TIA) array. To achieve a large bandwidth density, high energy efficiency and an error-free operation, TIAs should provide high gain and bandwidth, low noise, and a compact footprint, while offering low energy consumption. Inductive peaking is a common technique used to extend the bandwidth of amplifiers[23], however, on-chip inductors have a large footprint and could significantly decrease the system bandwidth density. Furthermore, undesired magnetic coupling between inductors of different channels could introduce inter-channel crosstalk in a WDM system.

Equalization systems such as decision feedback equalizer (DFE) or feedforward equalizer (FFE) can be employed in optical receivers to mitigate inter-symbol interference (ISI) and signal distortion[24] at the cost of decreased system energy efficiency and bandwidth density especially at higher data-rates. Forward error correction (FEC)[25] relaxes analog SNR requirements at the cost of lower net throughput (due to added redundant data), additional power consumption, increased latency and reduced bandwidth density.

Here, we report the demonstration of a monolithically integrated, electronic-photonic co-designed WDM receiver on a photonics-enabled 45nm CMOS SOI platform that achieves an aggregate data-rate of 1.024 Tb/s across 32 wavelength channels, concurrently modulated using NRZ format at 32 Gb/s/channel with a bit-error-rate (BER) below $10^{-12}$. Inverse designed grating couplers with a coupling loss of 4.5 dB and a -1-dB bandwidth of 25 nm were used, allowing for a wideband operation required for error-free detection of 32 modulated channels using 16 optical carriers with carrier-to-carrier spacing of 200 GHz. Monolithic integration of photodiodes with TIAs minimizes interface parasitic components and relaxes the gain–bandwidth–noise trade-off in TIA design, which together with low-loss demultiplexing and high bandwidth and low-noise electronics, enables error-free operation at 32 Gb/s/channel without use of any equalization or error correction schemes. Besides elimination of backend digital processing, error correction and equalization, implementation of capacitive tuning of micro-ring devices (with a zero static power consumption) and sharing the control electronics in a sequential wavelength locking scheme leads to a record energy efficiency of 71 fJ/b, which includes the energy consumption of the electronic circuitry and the wavelength tuning and locking of photonic devices.

Benefiting from resonance based photonic devices with autonomous wavelength locking and compact inductor-less trans-impedance amplifiers, the implemented optical receiver chip achieves a record bandwidth density of 4 Tb/s/mm$^2$.

To improve reliability, for cases that a tuning range larger than that of capacitive modulators is needed to overcome the unexpected effect of process variations, a hybrid capacitive-thermal tuning approach is utilized for micro-ring devices to optimize tuning efficiency while minimizing the energy consumption. The implemented energy-efficient and compact monolithic optical receiver offers a scalable and reliable solution for the optical interconnects of future datacenters.

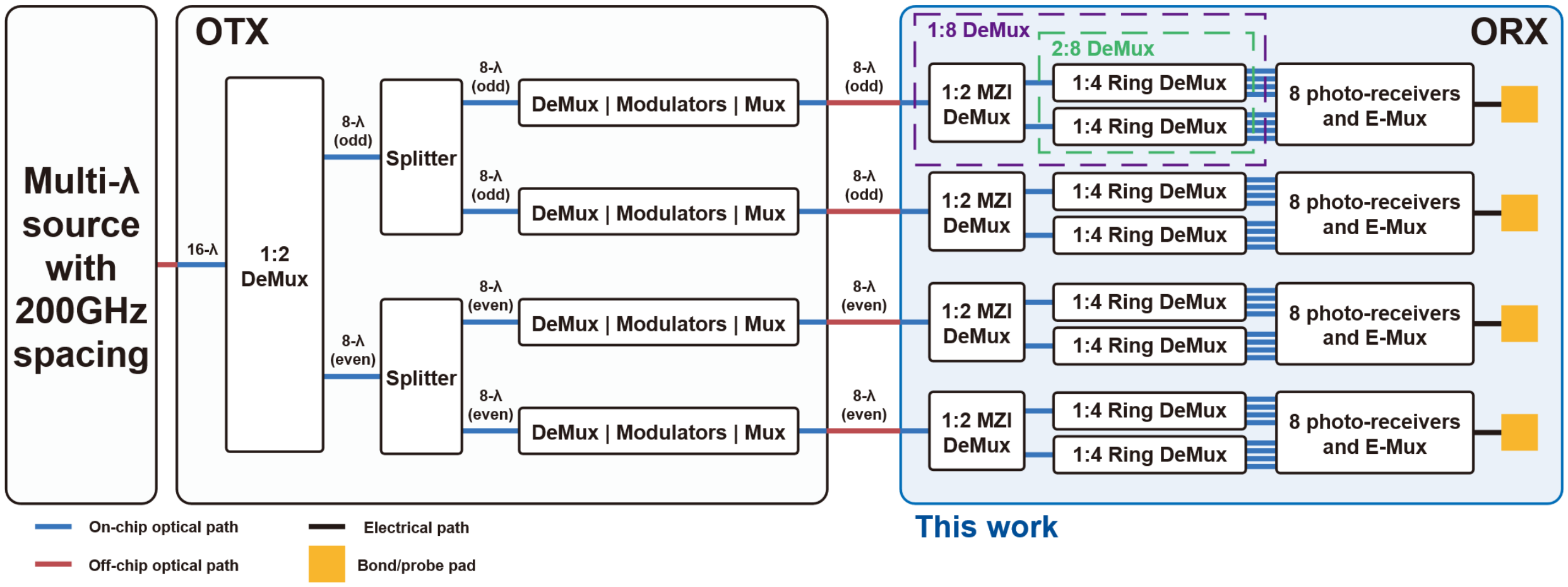

**Fig. 1 | Overview of the 1Tb/s receiver architecture.** The optical transmitter (OTX) separates the input 16 carriers into even and odd wavelength channels and equally splits each into two optical signals and independently modulates the resulting 32 channels and couples out the modulated carriers into 4 single-mode optical fibers. The fibers deliver the modulated carriers to the optical receiver (ORX) chip, where they are demultiplexed using a 1:2 MZI based stage followed a 2:8 stage (formed using two parallel 1:4 ring resonator based DeMux stages). The outputs of 2:8 DeMux stages are photo-detected, amplified, and routed off-chip for monitoring. MZI: Mach-Zehnder interferometer, Mux: multiplexer, DeMux: de-multiplexer, E-Mux: electrical multiplexer

# Results

## System architecture

Figure 1 shows the block diagram of the target optical link utilizing the implemented receiver chip. The output of a multi-wavelength source (e.g. as optical frequency comb or a laser array) with 16 modes (carriers) uniformly spaced on a 200 GHz grid is coupled to the optical transmitter (OTX). In this configuration, carriers are demultiplexed into even and odd wavelength channels and each carrier is split into two separate waveguides with equal power. The resulting 32 optical signals are individually modulated with independent data streams, each at 32 Gb/s, coupled out of the OTX using four single mode fibers (two for even wavelength channels and two for odd wavelength channels) and coupled into the four inputs of the 32-channel optical receiver (ORX) chip (this work). In this case, each fiber delivers eight carriers with a 400 GHz carrier-to-carrier spacing (either odd or even wavelength channels) centered near 1550 nm. Upon coupling to the ORX chip, first, a 1:2 MZI based DeMux is used to demultiplex each fiber output (with 8 carriers) into two branches, each with 4 carriers with an 800 GHz carrier-to-carrier spacing. This wider spacing helps reduce crosstalk requirements for the subsequent 1:4 ring resonator based DeMux stage, improving overall signal integrity. The 8 outputs of every two 1:4 DeMux stages (equivalently forming a 2:8

DeMux) are routed to an 8-channel photo-receiver array detecting and demodulating eight carriers. The electrical outputs of each 8-channel photo-receiver array are multiplexed using an 8:1 electrical multiplexer (E-Mux) and routed off-chip for monitoring.

## Optical Demultiplexer

The 32 channel NRZ modulated carriers are delivered to the optical receiver chip using four single-mode fibers and processed using four parallel on-chip 1:8 DeMux subsystems followed by an array of photo-receivers. Figure 2a shows the structure of the on-chip 1:8 DeMux subsystem. One of four SMFs, guiding 8 modulated carriers with a 400 GHz carrier-to-carrier spacing in optical C band, is coupled into the chip using an inverse designed grating couplers (with a coupling loss of 4.5 dB) and routed to the input of a 1:2 Mach-Zehnder interferometer (MZI) based DeMux using a nanophotonic waveguide. The 1:2 MZI-based DeMux, operating as a de-interleaver, separates the even and odd wavelength channels on its two output waveguides. The MZI output waveguides, each containing 4 carriers with an 800 GHz carrier-to-carrier spacing, sever as the input bus waveguides of a 2:8 DeMux, where 3 micro-ring resonators are placed on each bus waveguide to select 3 carriers (on their drop ports) while the last carrier remains on the bus waveguide output. The 8 outputs of the 2:8 micro-ring resonator based DeMux (6 on the drop ports of the rings and two at the output of the bus waveguides) are routed to an array of photo-receiver for detection.

Figure 2b shows the structure of the 1:2 MZI based DeMux (de-interleaver), where two wideband 50% directional couplers[26] are used to form the MZI. Meandered waveguides are used to set the delay difference between the two arms of the MZI to 1.25 ps. Thermal phased shifter are used to tune the response of the MZI. Figure 2c shows the measured response of the MZI for different voltages applied across the heater. The MZI heater achieves a $\pi$ phase shift power efficiency, $P_\pi$, of 10 mW.

Besides reducing the electronic-photonic interconnect parasitics, monolithic integration of the ORX chip enables implementation of capacitive optical phase shifters[22,27]. Figure 2d shows the structure of the implemented capacitive modulator in GlobalFoundries 45CLO CMOS process, where an optical waveguide is formed by vertically stacking silicon and polysilicon layers (the latter is typically used as the gate in MOSFET devices), separated by a thin $SiO_2$ layer. While this thin oxide layer does not directly interact with the optical mode, it acts as an insulating barrier to DC current when a voltage is applied across the polysilicon-oxide-silicon capacitor[22], eliminating the static power consumption. By applying a DC voltage across polysilicon-oxide-silicon capacitor, the charge distribution within the optical waveguide (overlapping with the optical mode) is adjusted, resulting in a change in the effective refractive index, thereby enabling optical phase adjustment.

Figure 2e shows the structure of the capacitively-tuned micro-ring resonator used within the 2:8 DeMux. The length of the through and drop coupling regions is 8 $\mu$m. The micro-ring, including both coupling regions are formed using a stack of polysilicon-oxide-silicon layers. The refractive index of the entire ring, except for the coupling regions, can be capacitively tuned. For all 3 port waveguides of the micro-ring, the polysilicon is tapered to gradually transition silicon waveguides to polysilicon-oxide-silicon waveguides and back. Although capacitive phase shifters provide the benefit of zero static power consumption, their wavelength tuning range is less than that of thermal phase shifters and while typically the capacitive tuning range is sufficient for normal operation of the micro-ring resonators, to increase the reliability and ensure correct operations for corner cases that the capacitive tuning range is insufficient, a hybrid tuning mechanism combining both thermal and capacitive control was implemented for each ring element, where slightly doped silicon resistors, positioned within the ring section, act as integrated heaters. Undercut trenches available in the 45CLO process are incorporated around the micro-rings for enhanced thermal tuning efficiency and minimized thermal crosstalk. Figure 2f shows the capacitive and thermal tuning response of the micro-ring resonator for different voltages across the capacitive section and the heater. The ring resonator heater provides a tuning efficiency of 125 pm/mW. Note that the heaters and capacitive phase shifters induce a redshift and a blueshift in the resonance wavelength of the ring, respectively. The maximum tuning range of capacitive phase shifters is about 0.7 nm.

The block diagram of the optical receiver chip is shown in Fig. 3a. Four input optical signals are coupled into the chip using an array of four inverse design grating couplers, each routing one of input optical signals to one of the four subsystems. Each subsystem consists of a 1:8 optical DeMux, an array of 1:8 photo-receivers, an 8:1 electrical multiplexer used for output monitoring and individual sensing and actuation units for each MZI and micro-ring resonator. A central monitoring and control unit is used to wavelength lock each MZI and ring resonator within the four subsystems to the corresponding target wavelength.

To compensate for undesired changes in the effective refractive index caused by temperature fluctuations or fabrication process variations, MZI and micro-ring devices are sequentially selected, their status (i.e. the location of the notch wavelength with respect to the wavelength of the target carrier) are monitored and the optical phase shifters within the MZIs or ring resonators (capacitive or thermal) are adjusted in a feedback loop. Figure 3b depicts the block diagram of the ring memory and actuation (RMA) unit, which consists of a capacitive actuation channel and a thermal actuation channel. The design of the memory and actuation channels follows our prior work[28]. Within the capacitive actuation channel, a 5-bit up/down counter, serving as a memory unit, is used to adjust the voltage across the corresponding capacitive phase shifter using a level-shifter followed by digital-to-analog converter (DAC). The electronic circuit in the thermal actuation channel is similar to that of the capacitive actuation channel except that it utilizes a 7-bit shift register as the input memory and also a heater driver at its output. Note that the number of bits in the counter and shift register is set based on the actuation gain and required tuning resolution.

Figure 3c shows the block diagram of the MZI memory and actuation (MMA) unit, which follows the same design as the capacitive actuation channel of the RMA but with a 6-bit counter and an output heater driver. The block diagram of the central monitoring and control (CMC) unit is shown in Fig. 3d. When a micro-ring/MZI is selected by the CMC unit, the status of the micro-ring/MZI is monitored with a sniffer photodiode tapping 1% of its output and routing the resulting photocurrent to the CMC unit after it is converted to a voltage using a resistor. Depending on the detected difference between the wavelength of the notch and the wavelength of the target carrier, the CMC unit increments or decrements the counter and adjusts the resonance wavelength of the micro-ring/MZI, until its resonance wavelength reaches the target wavelength (set by a predefined reference voltage stored in on-chip memory), at which point, the counter stops counting and stores its value (corresponding the target resonance wavelength) and the CMC unit selects the next device for tuning and control.

Figure 3e illustrates the sequential tuning and locking process for both MZI and ring resonators. Note that in this measurement the data acquisition is slowed down to ease the illustration. As shown in Fig. 3f and 3g, upon completion of the alignment, the wavelength response of the MZIs and micro-ring resonators are aligned with the input wavelength grid across both odd- and even-indexed wavelength buses.

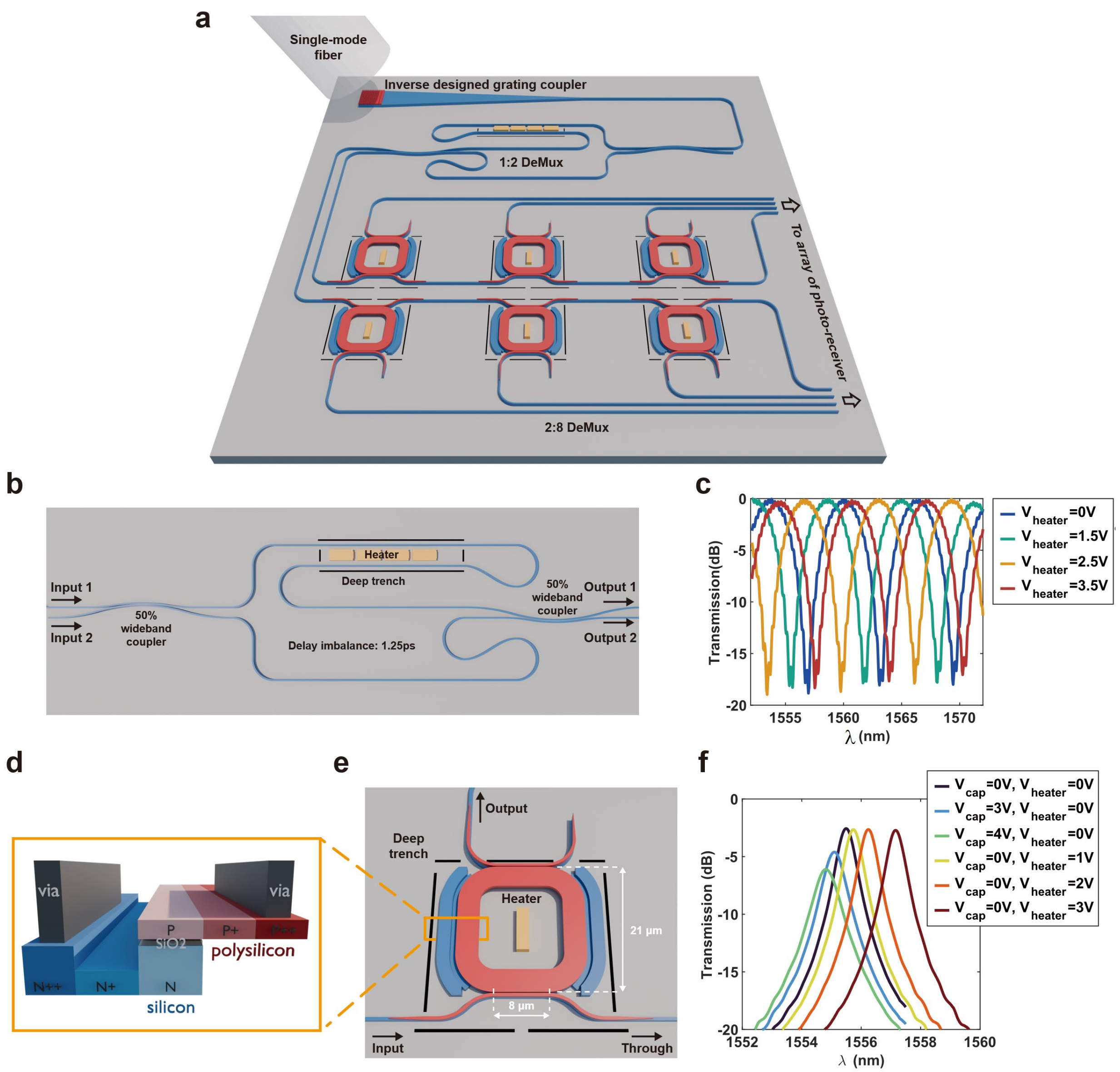

**Fig. 2 | Schematic of the DeMux subsystem. a,** Block diagram of the 1:8 optical DeMux subsystem, including the input inverse designed grating coupler, the 1:2 MZI based DeMux and 2:8 micro-ring resonator based DeMux. **b,** Structure of the 1:2 MZI based optical DeMux. **c,** MZI transmission response for different heater voltages. **d,** Structure of the capacitive phase shifter formed by vertically stacking polysilicon, thin oxide, and silicon layers. **e,** Structure of the capacitively tuned micro-ring resonator used to implement the 2:8 DeMux. **e,** Micro-ting resonator drop port response for different capacitive and heater voltages.

The DeMux units together consume an average static power of 10 mW to correct for the refractive index variations resulting from fabrication process variations, contributing 10 fJ/bit to the overall energy consumption of the receiver. The DeMux achieves a channel-to-channel isolation greater than 15 dB with an insertion loss of approximately 2 dB.

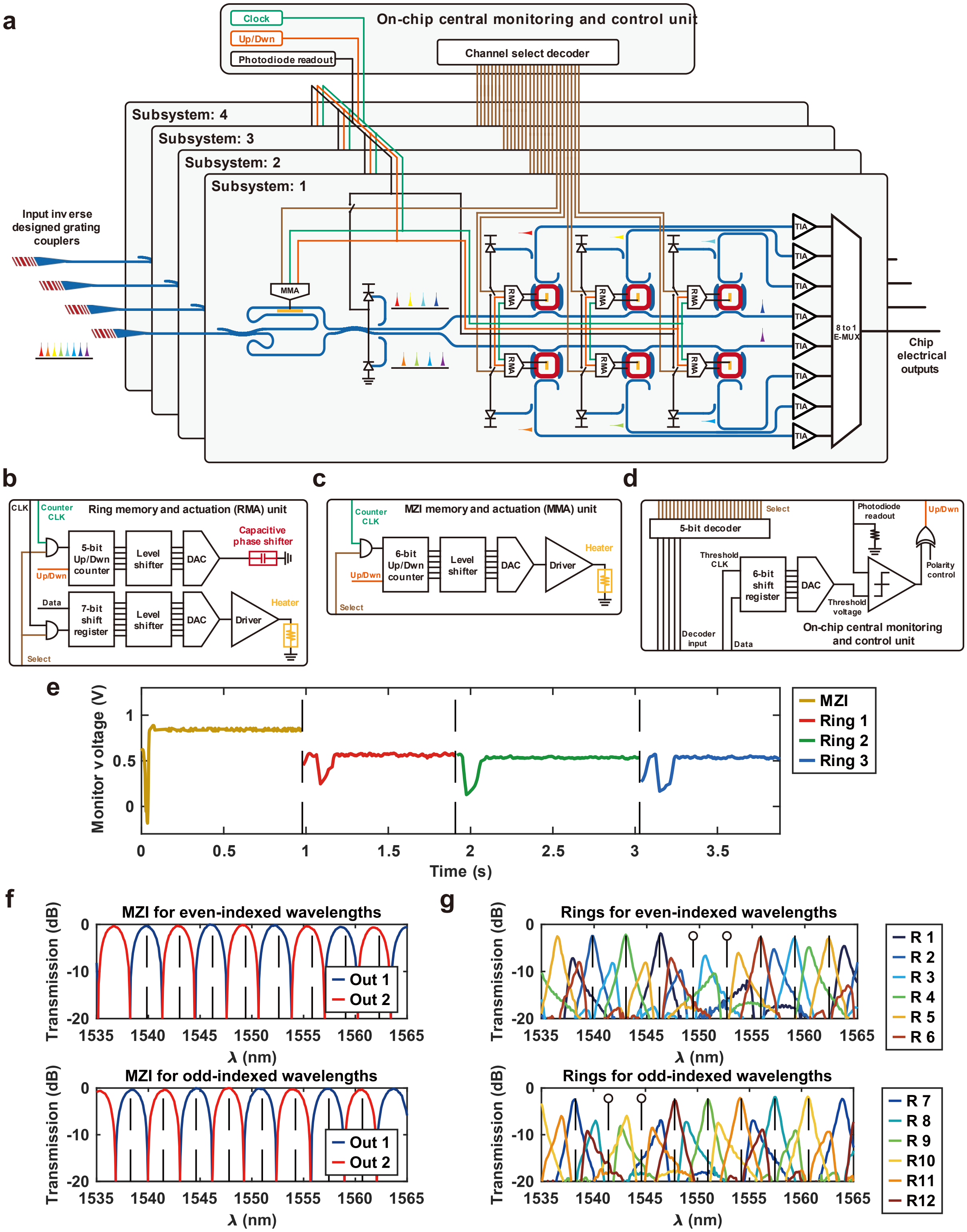

**Fig. 3 | Control and measurement results of the optical DeMux system. a,** Block diagram of the 4:32 optical DeMux at the input of the ORX chip, which consists of four 1:8 optical DeMux and the related memory, actuation, and control units. **b,** Block diagram of the ring memory and actuation (RMA) unit with two actuation channels for capacitive and thermal tuning. **c,** Block diagram of the MZI memory and actuation (MMA) unit used for thermal tuning of the MZI based 1:2 DeMux. **d,** Block diagram of the central monitoring and control (CMC) unit that sequentially selects and wavelength locks MZI and ring resonator devices. **e,**

Measured sequential wavelength locking of the MZI and ring resonators, where by individually enabling the control unit of each device, first, an open-loop sweep of the counter is performed to acquire and store the reference voltage corresponding to the target wavelength, and then, during the closed-loop control phase, the device is tuned to reach the target operating point. **f,** Wavelength response of the MZIs for even- and odd-indexed wavelengths after alignment. **g,** Wavelength response of the ring resonators for even and odd buses following wavelength alignment. The vertical dashed lines marked with "o" correspond to the wavelength of the carriers that are not selected by the 3 ring resonators on each bus (and remain on the bus as two of the eight DeMux outputs within each DeMux subsystem). Ri represents the i$^{th}$ micro-ring.

## Trans-impedance amplifier (TIA) array and readout circuit

In each subsystem, the outputs of the 1:8 DeMux are routed to an array of 8 photo-receivers. Within each photo-receiver (Fig. 4a), a SiGe photodiode with a responsivity of 0.9 A/W and bandwidth of 50 GHz converts the optical power of one of the DeMux outputs to a current and the resulting photo-current is amplified and converted to a voltage using a TIA. The TIA is designed to achieve error-free operation at 32 Gb/s/channel while maximizing the energy efficiency and bandwidth density. Figure 4b shows the schematic of the implemented inductor-less 3-stage Cherry-Hopper inverter-based TIA that utilizes transconductance to capacitance ratio scaling benefits and operates at lower supply voltage without having a large headroom constraint. The TIA achieves a simulated transimpedance gain of 74 dBΩ, a -3 dB bandwidth of 15 GHz and a total integrated input-referred current noise of 1.3 μA. The inductor-less compact TIA is implemented within a footprint of 43 um × 12 um corresponding to a bandwidth density of about 62 Tb/s/mm$^2$.

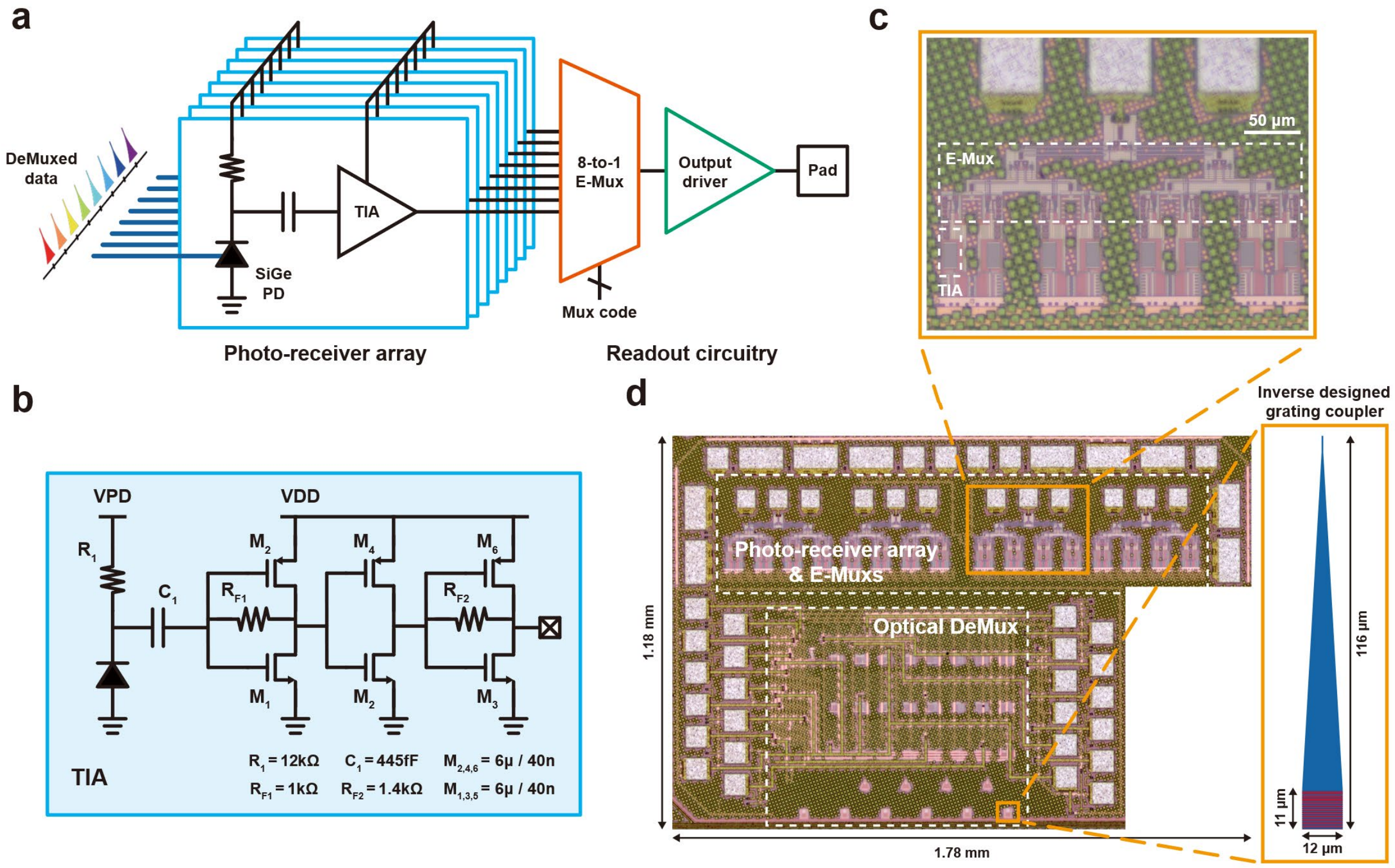

**Fig. 4 | Photo-receiver array design and chip micrograph | a,** Architecture of the photo-receiver array within one subsystem, which consists of 8 SiGe photodiodes (PDs) followed by 8 trains-impedance amplifiers. An 8:1 electrical multiplexer (E-Mux) is used to select one TIA output at a time. The output driver is used to drive measurement instruments. **b,** Schematic of the inductor-less 3-stage Cherry-Hopper TIA. **c,** Micrograph of the photo-receiver array**. d,** Micrograph of the implemented 1.024 Tb/s receiver chip.

To ease the system measurements, an on-chip 8:1 electrical multiplexer, E-Mux, is used within each subsystem to sequentially select the outputs of the 8 TIAs within the photo-receiver array for measurements without significantly degrading the performance. The E-Mux is followed by an output driver to drive the load presented by measurement instruments. More details on the design and characterization of the TIA, E-Mux and output driver are included in the Methods section and Extended Data Fig. 1.

Figure 4c shows the micrograph of the photo-receiver array, the E-Mux and the output driver. The micrograph of the implemented optical receiver chip is shown in Fig. 4d. The chip was fabricated using the GlobalFoundries 45CLO CMOS SOI process within a footprint of 2.1mm$^2$.

## System measurement

Figure 5a shows the measurement setup used to characterize the performance of the implemented optical receiver chip. Since on-chip subsystems are independent, the four on-chip subsystems were measured sequentially, while all modulated carriers for each subsystem are present concurrently. For each subsystem measurement, the input consists of eight modulated frequency tones spaced 400 GHz apart. These tones correspond to either the odd indexed wavelengths ($\lambda_1$, $\lambda_3$, …, $\lambda_{15}$) for subsystems 1 and 2 measurements (through bus 1 and bus 2), or the even indexed wavelengths ($\lambda_2$, $\lambda_4$, …, $\lambda_{16}$) for subsystems 3 and 4 measurements (through bus 3 and bus 4).

To generate the input modulated optical signals for subsystems 1 and 2, four outputs of a Thorlabs PRO8000 laser bank (with 4 distributed feedback laser modules) are combined representing carriers with wavelengths $\lambda_1$, $\lambda_5$, $\lambda_9$ and $\lambda_{13}$, with 800 GHz carrier-to-carrier spacing, within a single fiber (Fiber 1). Similarly, four outputs of another Thorlabs PRO8000 laser bank are combined representing carriers with wavelengths $\lambda_3$, $\lambda_7$, $\lambda_{11}$ and $\lambda_{15}$ (also with 800 GHz carrier-to-carrier spacing), within another fiber (Fiber 2). To fully capture the impact of on-chip optical crosstalk, which can significantly degrade system performance, the two optical fibers (i.e. Fiber 1 and Fiber 2), each carrying four wavelengths spaced 800

GHz apart, are separately modulated using two different 32Gb/s pseudo-random binary sequences of order 7 (PRBS-7) generated using two independent output channels of a Micram DAC10002 arbitrary waveform generator, and are then interleaved. The resulting eight modulated carriers with wavelengths $\lambda_1, \lambda_3, \ldots, \lambda_{15}$ that are transmitted in a single fiber are amplified using an erbium-doped fiber amplifier (EDFA) and coupled into subsystem 1 (or 2). Each of the eight carriers is modulated with a different data stream compared to its neighboring carriers. In this case, if there is crosstalk between carriers with adjacent wavelengths, it will lead to data corruption and an increase in bit-error-rate (BER). Same measurement setup is used for characterization of subsystems 3 and 4 but four different lasers of the two laser banks were utilized to generate carriers at $\lambda_2, \lambda_4, \ldots, \lambda_{16}$.

Within each subsystem, the 1:8 optical DeMux is initialized using an off-chip micro-controller and is autonomously tuned to align to the International Telecommunication Union (ITU) wavelength grid using on-chip tuning and control loops. The detailed wavelength mapping is shown in Fig. 5b, where the channel number represents the order of the optical DeMux output waveguides routed to the TIA array. To ease the measurements, the number of output pads needed for readout of all channels is reduced by utilizing four on-chip 8-to-1 E-Muxs (i.e. one per subsystem) so that all 32 channels can be measured using 4 groups of ground-signal-ground (GSG) radio frequency pads. A DCA-J 86100C sampling scope and an SHF 11100A error analyzer were used to monitor the eye-diagram and BER of the electrical outputs of the optical receiver, respectively. Figure 5c shows the measured eye diagrams for all 32 channels, where the uniformity shows the potential to scale the receiver chip to an even higher aggregate data-rates by integrating more channels. The measured BER is shown in Fig. 5d, where a BER < $10^{-12}$ at 32 Gb/s were measured for all channels indicating a uniform performance of all photonic and electronic components across entire chip. Note that for eye diagram and BER measurements, all modulated carriers of each subsystem are concurrently present. Given that the bit-stream for neighboring channels (i.e. modulated carriers with one carrier-to-carrier spacing lower or higher than the carrier being measured) are different, the effect of optical and electrical crosstalk between adjacent channels, while small, is present in all measurements. The receiver chip demonstrates a measured OMA sensitivity of up to -8.1 dBm (Fig. 5e). Figure 5f shows the measured bathtub curve indicating that the receiver maintains a 22% unit interval (UI) timing margin while operating at BER < $10^{-12}$. The total energy consumption, including the optical DeMux tuning and locking, and the array of TIAs, is about 71 fJ/bit. The core components, including the photonic devices, tuning electronics, and TIAs, occupy an area of about 0.25 mm², resulting in an areal bandwidth density of more than 4 Tb/s/mm².

## Discussion

In this work, we have demonstrated a monolithically integrated electronic-photonic WDM receiver, fabricated using a 45 nm CMOS SOI platform, that achieves an aggregate data-rate of 1.024 Tb/s across 32 wavelength channels. Each channel operates at 32 Gb/s with a BER of less than $10^{-12}$, enabled by a compact, energy-efficient architecture that delivers a record-low energy efficiency of 71 fJ/bit and a bandwidth density of 4 Tb/s/mm² while eliminating the need to equalization and error-correction.

The implementation of a hybrid tuning scheme, which combines capacitive phase shifters with zero-static-power consumption and localized thermal phase shifters, balances tuning range, thermal stability, and power consumption, enabling low-energy yet robust wavelength alignment across all channels. The built-in autonomous wavelength tuning and locking ensures stable optical alignment using on-chip feedback, eliminating the need for external calibration mechanisms. Arrays of inductor-less, inverter-based Cherry-Hooper trans-impedance amplifiers optimized for low-voltage operation with a gain of 74 dBΩ, a -3 dB bandwidth of 15 GHz, a small footprint and a high energy efficiency were utilized.

The implemented chip is an energy-efficient scalable high-density solution for next-generation data center interconnects. Extended Data Table 1 compares the performance of the implemented optical receiver chip with that of other works.

In this work, the combination of monolithic integration, energy-efficient tuning, compact design, low-loss photonics and power-efficient and low-noise electronics (leading to elimination of the need for DSP, error correction and equalization) make the implemented highly energy efficient and compact optical receiver suited for deployment in future hyperscale datacenters and HPC systems.

# Methods

## Chip fabrication

The optical receiver chip was monolithically integrated using the GlobalFoundries 45CLO CMOS-SOI silicon photonic process, which offers an $f_T$ of up to 280 GHz[29].

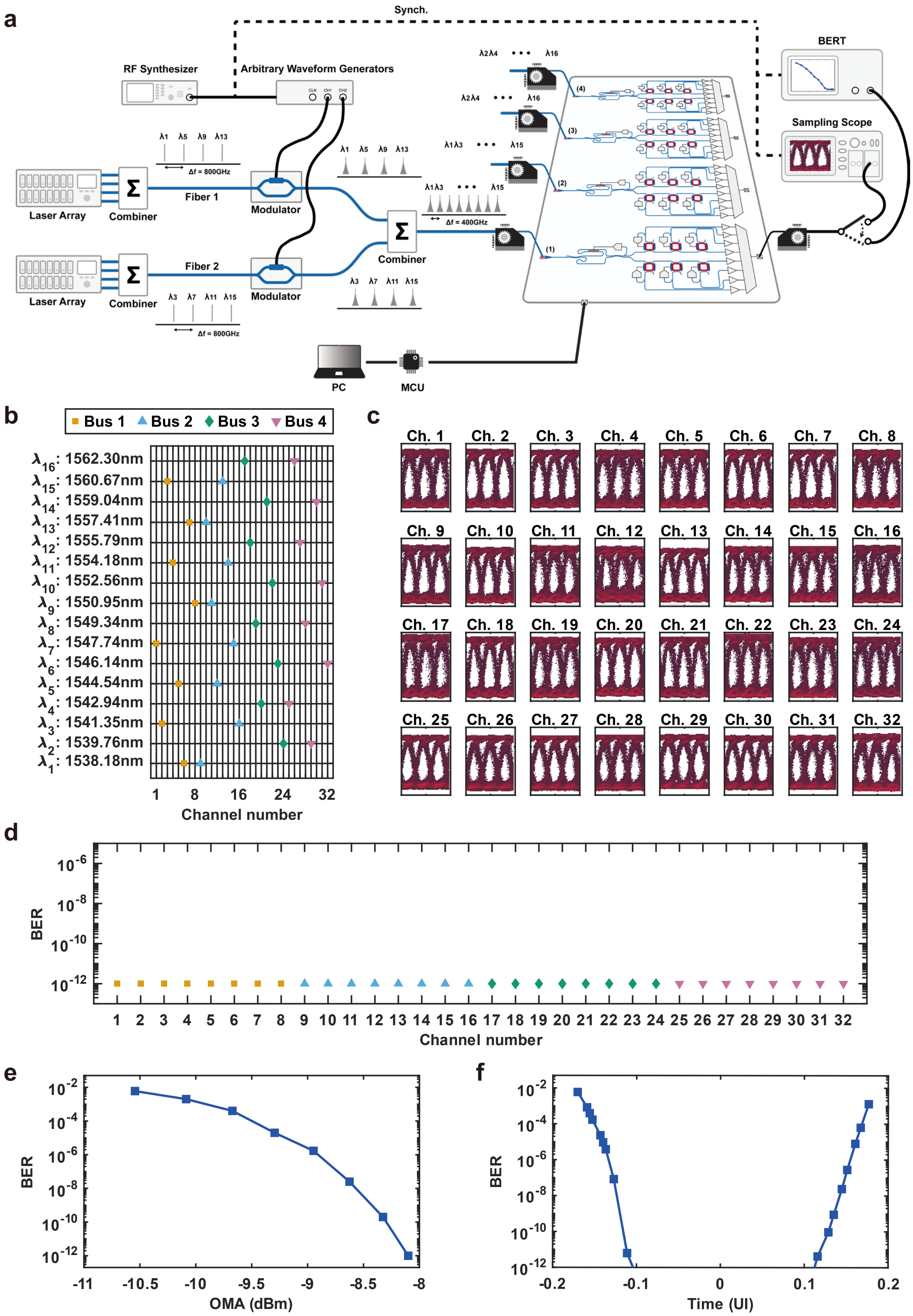

**Fig. 5 | Measurements of the 32-channel optical receiver chip. a,** Measurement setup including the input laser banks, modulation scheme ensuring different bit-streams for adjacent channels, monitoring and control scheme and output electrical eye diagram and BER characterization systems. **b,** Wavelength-

channel mapping spanning 24 nm of input optical bandwidth in C band. **c,** Measured eye diagrams and **d,** bit-error-rates of all 32 channels operating at 32 Gb/s/channel while all modulated carriers of subsystems are concurrently present. **e,** Measured BER versus received optical power at the photodiode input for a typical channel. **f,** Measured receiver bathtub curve.

The design incorporates 500nm-wide single-mode photonic waveguides with an optical loss of about 1.4 dB/cm at around 1550 nm, inverse designed grating couplers with a coupling loss of about 4.5 dB and a -1-dB bandwidth of about 25 nm, and SiGe photodiodes with a responsivity of approximately 0.9 A/W and a bandwidth of about 50 GHz.

## Inverse designed grating couplers

The grating couplers utilized in this system are designed using the photonic inverse design method based on a variation of the software package (SPINS)[30,31]. Three important considerations are made in the photonic inverse design process for the grating coupler design used in this system. First, the polysilicon layer on top of the silicon slab is chosen to be optimized to break the z-symmetry of the grating coupler design. This single layer optimization minimizes the performance variation induced by the layer-to-layer misalignment. We then frame the objective function to target broadband wavelength response to meet the wide frequency bandwidth of the multi-carrier source utilized for multiplexed channels. The objective function with the utilization of finite difference frequency domain (FDFD) solver can be described as

$$f_{obj}(\epsilon) = \sum_i \left(1 - \left|\mathcal{L}\left(E_i(\epsilon)\right)\right|\right)^2, \tag{1}$$

where $\left|\mathcal{L}\left(E_i(\epsilon)\right)\right|$ is the overlap integral of the vectorized electric field ($E$) at optical frequencies ($\omega_i$) with the waveguide mode. Lastly, the optimization is constrained by the fabrication design rules (e.g. minimum feature and gap sizes) of the GlobalFoundries 45CLO process.

## TIA design, electrical multiplexer, and driver

In Fig. 4b, a large resistor, $R_1$, is used to reverse bias the photodiode allowing most of the ac photocurrent to flow into the TIA, while the DC bias is isolated from TIA input by a large dc blocking capacitor, $C_1$. The low frequency corner of the photo-receiver is carefully set after balancing the tradeoff between the bottom plate parasitic capacitance of the Metal-Oxide-Metal (MOM) capacitor and the low frequency content of the incoming data stream.

The input-referred current noise of the TIA is a key factor for achieving low bit-error-rate. The input referred current noise of the TIA in Fig. 4b can be calculated as[32]

$$\overline{I_{n,\text{in}}^2} \approx \frac{4kT}{R_F} + \frac{4kT\gamma}{g_m R_F^2}\left|1 + sR_F C_{\text{tot,in}}\right|^2, \tag{2}$$

where $k$, $T$, $R_F$, $\gamma$, $g_m$ and $C_{\text{tot,in}}$ are the Boltzmann constant, the temperature in Kelvins, feedback resistor, the channel-noise factor, the equivalent transconductance of the first stage of the amplifier and the total capacitance at the TIA input, respectively.

For a given target bandwidth, the small parasitic capacitance of the photodiode-TIA interface (enabled by monolithic integration) allows for use of a larger feedback resistance, $R_F$, which results is a higher trans-impedance gain and, as Eqn. 2 indicates, a lower input-referred current noise. The simulated trans-impedance gain and the input-referred current noise power spectral density of the TIA is shown in Extended Data Fig. 1a.

Extended Data Fig. 1b shows the schematic of the implemented 8:1 E-Mux, which is formed using a 3-layer binary tree of 2:1 E-Mux units. During the readout process, a specific E-Mux code activates the 2:1 E-Mux units along the selected signal path, while the rest of the 2:1 E-Mux units are turned off to minimize the crosstalk. The schematic of the 2:1 E-Mux is shown in Extended Data Fig. 1c. The schematic of the voltage mode CMOS driver used after the 1:8 E-Mux to drive the off-chip measurement instruments is depicted in Extended Data Fig. 1d.

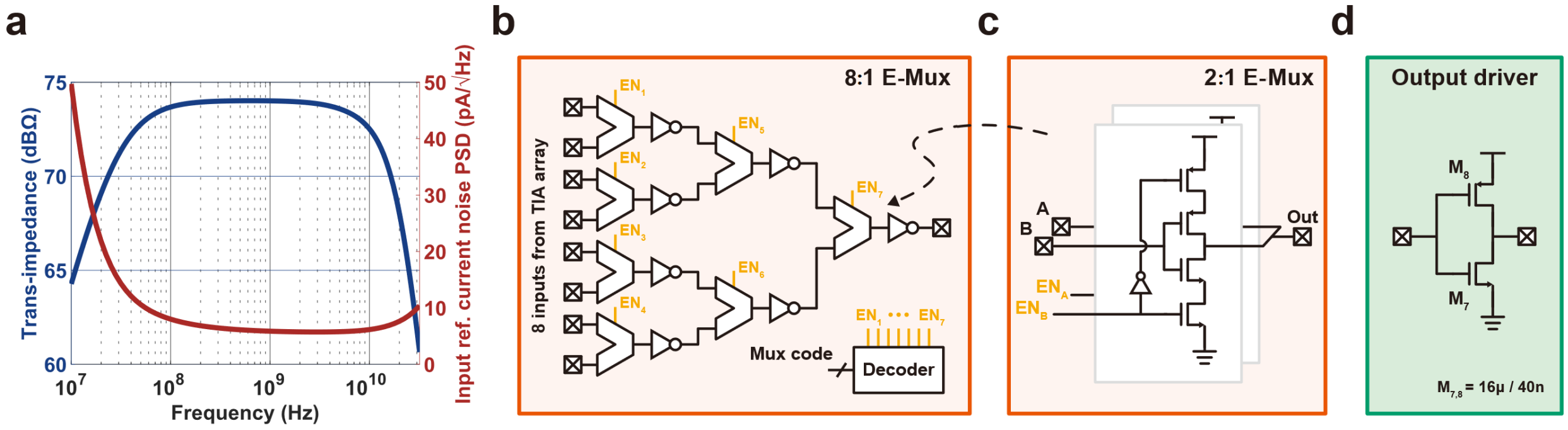

**Extended Data Fig. 1 | Photo-receiver electronics. a,** Simulated trans-impedance and the power spectral density of the input-referred current noise of the TIA. **b,** Schematic of the 8:1 E-Mux formed using 3 multiplexing binary tree layers of 2:1 E-Mux units. **c,** Schematic of the 2:1 E-Mux. **d,** Schematic of the output driver.

## Acknowledgement

This work was supported by Defense Advanced Research Projects Agency PIPES program under contract HR0011-19-2-0016.

## Author contributions

A.P., Z.Y., J.V. and F.A. conceived the project idea. A.P. and Z.Y. contributed equally to this work and led the overall chip and PCB design and conducted full system measurements with assistance from A.S. G.H.A. designed and characterized the inverse-designed grating couplers. J.V. directed and supervised the design, implementation and characterization of inverse designed devices. All authors contributed to writing the manuscript. F.A. directed and supervised the project.

| | [23] | [33] | [34] | [35] | This work |
|---|---|---|---|---|---|
| Integration approach | Hybrid | Hybrid | Monolithic | Hybrid | Monolithic |
| PIC / EIC Technology | 45nm SiPh/ 7nm FinFET | In-house/ 28nm CMOS | 45nm CMOS SOI-Photonic | 180nm SiPh/ 28nm CMOS | 45nm CMOS SOI-Photonic |
| WDM Frequency Spacing [GHz] | 250 | 200 | 400 | 500 | 200 |
| DeMux locking | N/A | Integrated | Integrated | Integrated | Integrated |
| DeMux tuning | N/A | Thermal | Thermal | Thermal | Capacitive and thermal |
| Channel Number | 1 fiber x 7 λ | 1 fiber x 8 λ | 1 fiber x 8 λ | 1 fiber x 4 λ[1] | 4 fibers x 8 λ |
| Concurrent measurement | Yes | Yes | Yes | No | **Yes** |
| NRZ aggregate data-rate (Gb/s) | 350 | 256 | 256 | 200[2] | **1024** |
| BER | 1e-12 | 1e-12 | 1e-15 | 1e-12[1] | 1e-12 |
| Receiver energy efficiency (pJ/bit) | 0.96 | 3.80 | 1.03 | 1.34 | **0.06** |
| Tuning energy efficiency (pJ/bit) | N/A | N/A | 0.1 | N/A | **0.01** |
| Total energy efficiency (pJ/bit) | >0.96[3] | >3.80[3] | 1.13 | >1.34[3] | **0.07** |
| Error correction | No | No | No | Yes | **No** |
| Equalization | Yes | Yes | Yes | Yes (off-chip) | **No** |
| Receiver bandwidth density (Tb/s/mm$^2$) | 1.61 (electronics only) | N/A | N/A | N/A | **4 Tb/s/mm$^2$** (electronics & photonics) |
| Sensitivity (OMA dBm) | -11.4 | -10.1 | -14.2 | -8.4 | -8.1 |

[1]Measured one carrier at a time
[2]Based on 50 Gb/s NRZ per-channel for hybrid-integrated system.
[3]Excluding thermal tuning

**Extended Data Table. 1 | Comparison with the state-of-the-art works.**

## Disclosures

The authors declare no conflicts of interest.

## Data availability

Data underlying the results presented in this paper are not publicly available at this time but may be obtained from the authors upon request.

## References

1. Daudlin, S., et al. Integrated Kerr Comb Link with Multi-Channel DWDM Silicon Photonic Receiver. in *Proc. Conference on Lasers and Electro-Optics (CLEO)* 2022 STu5G.3 (Optica Publishing Group, 2022).

2. Li, C., et al., Hybrid WDM-MDM transmitter with an integrated Si modulator array and a micro-resonator comb source. *Optics Express* **29**, 39847-39858 (2021).

3. Liu, A., et al., Wavelength division multiplexing based photonic integrated circuits on silicon-on-insulator platform. *IEEE Journal of Selected Topics in Quantum Electronics* ***16***, 23-32 (2009).

4. Omirzakhov, K., et al. Monolithically Integrated Sub-63 fJ/b 8-Channel 256Gb/s Optical Transmitter with Autonomous Wavelength Locking in 45nm CMOS SOI. in *IEEE International Solid-State Circuits Conference (ISSCC),* 12.1 (2024).

5. Pfeifle, J., et al., Coherent terabit communications with microresonator Kerr frequency combs. *Nature photonics* **8,** 375-380 (2014).

6. Rizzo, A., et al., Massively scalable Kerr comb-driven silicon photonic link. *Nature Photonics* **17**, 781-790 (2023).

7. Sharma, J., et al., Silicon photonic microring-based 4× 112 Gb/s WDM transmitter with photocurrent-based thermal control in 28-nm CMOS. *IEEE Journal of Solid-State Circuits* **57**, 1187-1198 (2021).

8. Shu, H., et al., Microcomb-driven silicon photonic systems. *Nature* **605**, 457-463 (2022).

9. Wade, M., et al. An error-free 1 Tbps WDM optical I/O chiplet and multi-wavelength multi-port laser. in *Proc. Optical Fiber Communication Conference (OFC)* 2021 F3C.6 (Optica Publishing Group 2021).

10. Rakowski, M., et al. 22.5 A 4× 20Gb/s WDM ring-based hybrid CMOS silicon photonics transceiver. in *IEEE International Solid-State Circuits Conference (ISSCC), 22.5 (2015).*

11. Yu, H., et al. 800 Gbps fully integrated silicon photonics transmitter for data center applications. in *Proc. Optical Fiber Communication Conference (OFC)* 2022 M2D.7 (Optica Publishing Group 2022).

12. Sun, C., et al., A 45 nm CMOS-SOI monolithic photonics platform with bit-statistics-based resonant microring thermal tuning. *IEEE Journal of Solid-State Circuits* **51**, 893-907 (2016).

13. Siew, S.Y., et al., Review of silicon photonics technology and platform development. *Journal of Lightwave Technology* **39**, 4374-4389 (2021).

14. Wang, L., et al., Athermal arrayed waveguide gratings in silicon-on-insulator by overlaying a polymer cladding on narrowed arrayed waveguides. *Applied optics* **51**, 1251-1256 (2012).

15. Lu, Z., et al., Performance prediction for silicon photonics integrated circuits with layout-dependent correlated manufacturing variability. *Optics express* **25**, 9712-9733 (2017).

16. Xing, Y., et al., Capturing the effects of spatial process variations in silicon photonic circuits. *ACS Photonics* **10**, 928-944 (2022).

17. Akiyama, T., et al., Cascaded AMZ triplets: a class of demultiplexers having a monitor and control scheme enabling dense WDM on Si nano-waveguide PICs with ultralow crosstalk and high spectral efficiency. *Optics express* **29**, 7966-7985 (2021).

18. Padmaraju, K., et al. Wavelength locking of a WDM silicon microring demultiplexer using dithering signals. in *Proc. Optical Fiber Communication Conference (OFC)* 2014 Tu2E.4 (Optica Publishing Group 2014).

19. Rizzo, A., et al., Ultra-broadband interleaver for extreme wavelength scaling in silicon photonic links. *IEEE Photonics Technology Letters* **33**, 55-58 (2020).

20. Grillanda, S., et al., Wavelength locking of silicon photonics multiplexer for DML-based WDM transmitter. *Journal of Lightwave Technology* **35**, 607-614 (2016).

21. De, S., et al., CMOS-compatible photonic phase shifters with extremely low thermal crosstalk performance. *Journal of Lightwave Technology* **39**, 2113-2122 (2021).

22. Pirmoradi, A. and F. Aflatouni. Monolithically Integrated Autonomous Demultiplexers with Near Zero Power Consumption for Beyond Tb/s Links. in *Proc. Optical Fiber Communications Conference (OFC)* 2023 M3I.2 (Optica Publishing Group 2023).

23. Raj, M., et al. A O. 96pJ/b 7× 50Gb/s-per-Fiber WDM Receiver with Stacked 7nm CMOS and 45nm Silicon Photonic Dies. in *IEEE International Solid-State Circuits Conference (ISSCC),* 12.1 (2023).

24. H. Li, C. -M. Hsu, J. Sharma, J. Jaussi and G. Balamurugan, A 100-Gb/s PAM-4 Optical Receiver With 2-Tap FFE and 2-Tap Direct-Feedback DFE in 28-nm CMOS. in *IEEE Journal of Solid-State Circuits* **57**, 44-53 (2022).

25. A. Udalcovs, R. Schatz, L. Wosinska, and P. Monti, Analysis of spectral and energy efficiency tradeoff in single-line rate WDM links. Journal of Lightwave **35**, 1847-1857 (2017).

26. A. Pirmoradi, H. Hao, K. Omirzakhov, A. J. Geers, and F. Aflatouni, A single chip 1.024 Tb/s silicon photonics PAM4 receiver. *arXiv preprint* arXiv:2507.12452 (2025).

27. A. Pirmoradi, J. Zang, K. Omirzakhov, Z. Yu, Y. Jin, S. B. Papp, and F. Aflatouni, Integrated multi-port multi-wavelength coherent optical source for beyond Tb/s optical links. *Nature Communications* **16**, 6387 (2025).

28. Omirzakhov, K., et al., Energy Efficient Monolithically Integrated 256 Gb/s Optical Transmitter With Autonomous Wavelength Stabilization in 45 nm CMOS SOI. *IEEE Journal of Solid-State Circuits* **60,** 2522-2531 (2025).

29. M. Rakowski et al., 45nm CMOS - Silicon Photonics Monolithic Technology (45CLO) for next-generation, low power and high speed optical interconnects. in *Optical Fiber Communication Conference (OFC)* 2020 T3H.3 (Optica Publishing Group 2020).

30. Su, L., et al., Fully-automated optimization of grating couplers. *Optics express* **26**, 4023-4034 (2018).

31. Su, L., et al., Nanophotonic inverse design with SPINS: Software architecture and practical considerations. *Applied Physics Reviews* **7**, 011407 (2020).

32. D. Li et al., A Low-Noise Design Technique for High-Speed CMOS Optical Receivers. in *IEEE Journal of Solid-State Circuits* **49**, 1437-1447 (2014).

33. Z. Xuan et al., A 256 Gbps Heterogeneously Integrated Silicon Photonic Microring-based DWDM Receiver Suitable for In-Package Optical I/O. *IEEE Symposium on VLSI Technology and Circuits,* C6-2 (2023).

34. Bhargava, P., et al. A 256Gbps Microring-Based WDM Transceiver with Error-Free Wide Temperature Operation for Co-Packaged Optical I/O Chiplets. in *IEEE Symposium on VLSI Technology and Circuits,* (2024).

35. J. Xue et al., "A 4×112 Gb/s Ultra-Compact Polarization-Insensitive Silicon Photonics WDM Receiver With CMOS TIA for Co-Packaged Optics and Optical I/O," in *Journal of Lightwave Technology* **42**, 6028-6035 (2024).